\documentclass[%
 reprint,
 superscriptaddress,
 amsmath,amssymb,
 aps,
]{revtex4-2}

\usepackage{graphicx}
\usepackage{dcolumn}
\usepackage{bm}
\usepackage{hyperref}
\usepackage{booktabs}
\usepackage{soul}
\usepackage{diagbox}
\usepackage{upgreek}
\usepackage{color}
\usepackage[defaultcolor=blue]{changes}
\usepackage[T1]{fontenc}
\usepackage[utf8]{inputenc}

\begin{document}

\preprint{APS/123-QED}

\title{In-plane optimal qubit operation with control fidelities exceeding 99 $\%$}

\author{Yu-Chen Zhou}
\thanks{These authors contributed equally to this work.}
\affiliation{Laboratory of Quantum Information, University of Science and Technology of China, Hefei, Anhui 230026, China}
\affiliation{Anhui Province Key Laboratory of Quantum Network, University of Science and Technology of China, Anhui 230026, China}
\affiliation{CAS Center for Excellence and Synergetic Innovation Center in Quantum Information and Quantum Physics, University of Science and Technology of China, Hefei, Anhui 230026, China}

\author{Zhenzhen Kong}
\thanks{These authors contributed equally to this work.}
\affiliation{Beijing Superstring Academy of Memory Technology, Beijing 100176, China}

\author{Rong-Long Ma}
\affiliation{Laboratory of Quantum Information, University of Science and Technology of China, Hefei, Anhui 230026, China}
\affiliation{Anhui Province Key Laboratory of Quantum Network, University of Science and Technology of China, Anhui 230026, China}
\affiliation{CAS Center for Excellence and Synergetic Innovation Center in Quantum Information and Quantum Physics, University of Science and Technology of China, Hefei, Anhui 230026, China}

\author{Hao-Tian Jiang}
\affiliation{Laboratory of Quantum Information, University of Science and Technology of China, Hefei, Anhui 230026, China}
\affiliation{Anhui Province Key Laboratory of Quantum Network, University of Science and Technology of China, Anhui 230026, China}
\affiliation{CAS Center for Excellence and Synergetic Innovation Center in Quantum Information and Quantum Physics, University of Science and Technology of China, Hefei, Anhui 230026, China}

\author{Ao-Ran Li}
\affiliation{Laboratory of Quantum Information, University of Science and Technology of China, Hefei, Anhui 230026, China}
\affiliation{Anhui Province Key Laboratory of Quantum Network, University of Science and Technology of China, Anhui 230026, China}
\affiliation{CAS Center for Excellence and Synergetic Innovation Center in Quantum Information and Quantum Physics, University of Science and Technology of China, Hefei, Anhui 230026, China}

\author{Yang Zhong}
\affiliation{Laboratory of Quantum Information, University of Science and Technology of China, Hefei, Anhui 230026, China}
\affiliation{Anhui Province Key Laboratory of Quantum Network, University of Science and Technology of China, Anhui 230026, China}
\affiliation{CAS Center for Excellence and Synergetic Innovation Center in Quantum Information and Quantum Physics, University of Science and Technology of China, Hefei, Anhui 230026, China}

\author{Zhi-Tao Wu}
\affiliation{Laboratory of Quantum Information, University of Science and Technology of China, Hefei, Anhui 230026, China}
\affiliation{Anhui Province Key Laboratory of Quantum Network, University of Science and Technology of China, Anhui 230026, China}
\affiliation{CAS Center for Excellence and Synergetic Innovation Center in Quantum Information and Quantum Physics, University of Science and Technology of China, Hefei, Anhui 230026, China}

\author{Rui Zheng}
\affiliation{Laboratory of Quantum Information, University of Science and Technology of China, Hefei, Anhui 230026, China}
\affiliation{Anhui Province Key Laboratory of Quantum Network, University of Science and Technology of China, Anhui 230026, China}
\affiliation{CAS Center for Excellence and Synergetic Innovation Center in Quantum Information and Quantum Physics, University of Science and Technology of China, Hefei, Anhui 230026, China}

\author{Gui-Lei Wang}
\email{guilei.wang@bjsamt.org.cn}
\affiliation{Beijing Superstring Academy of Memory Technology, Beijing 100176, China}
\affiliation{Hefei National Laboratory, Hefei 230088, China}

\author{Gang Cao}
\affiliation{Laboratory of Quantum Information, University of Science and Technology of China, Hefei, Anhui 230026, China}
\affiliation{Anhui Province Key Laboratory of Quantum Network, University of Science and Technology of China, Anhui 230026, China}
\affiliation{CAS Center for Excellence and Synergetic Innovation Center in Quantum Information and Quantum Physics, University of Science and Technology of China, Hefei, Anhui 230026, China}
\affiliation{Hefei National Laboratory, Hefei 230088, China}

\author{Hai-Ou Li}
\email{haiouli@ustc.edu.cn}
\affiliation{Laboratory of Quantum Information, University of Science and Technology of China, Hefei, Anhui 230026, China}
\affiliation{Anhui Province Key Laboratory of Quantum Network, University of Science and Technology of China, Anhui 230026, China}
\affiliation{CAS Center for Excellence and Synergetic Innovation Center in Quantum Information and Quantum Physics, University of Science and Technology of China, Hefei, Anhui 230026, China}
\affiliation{Hefei National Laboratory, Hefei 230088, China}

\author{Guo-Ping Guo}
\affiliation{Laboratory of Quantum Information, University of Science and Technology of China, Hefei, Anhui 230026, China}
\affiliation{Anhui Province Key Laboratory of Quantum Network, University of Science and Technology of China, Anhui 230026, China}
\affiliation{CAS Center for Excellence and Synergetic Innovation Center in Quantum Information and Quantum Physics, University of Science and Technology of China, Hefei, Anhui 230026, China}
\affiliation{Hefei National Laboratory, Hefei 230088, China}
\affiliation{Origin Quantum Computing Company Limited, Hefei, Anhui 230088, China}





\date{\today}
\begin{abstract}
Hole spin qubits based on semiconductor quantum dots are promising for building future large-scale quantum computers owing to their all-electrical manipulation. However, abundant physical mechanisms of valence band holes lead to anisotropic qubit properties. There is an opportunity to prolong the coherence time and achieve high-fidelity qubit manipulations. Here, we report a single-hole spin qubit in a planar germanium quantum dot and investigate its anisotropic susceptibility to charge noise under an in-plane magnetic field. By correlating the longitudinal spin-electric susceptibility with qubit coherence and control performance, we identify an optimal operating point where the sensitivity to charge noise is minimized. We find that optimizing the magnetic-field orientation reduces the spin-electric susceptibility, resulting in a fivefold enhancement of the Hahn-echo coherence time and a nearly tenfold suppression of control infidelity. At the optimal operating point, gate set tomography demonstrates the maximum gate fidelity of 99.82 $\%$. Our finding enhances the prospects of hole spin qubits for scalable quantum information processing.
\end{abstract}

\maketitle


\section{\label{sec:level1}INTRODUCTION}

Spin qubits \cite{lossQuantumComputationQuantum1998b} based on semiconductor quantum dots represent a promising platform for large-scale quantum computing\cite{zhangSemiconductorQuantumComputation2019c,stanoReviewPerformanceMetrics2022}, owing to their exceptional properties. These qubits offer advantages including extended coherence times \cite{veldhorstAddressableQuantumDot2014c} and small footprints \cite{neyensProbingSingleElectrons2024}, while simultaneously leveraging the well-established semiconductor manufacturing infrastructure for scalability \cite{george12SpinQubitArraysFabricated2025}. In particular, the hole in germanium has garnered significant attention owing to several favorable characteristics for encoding spin qubits \cite{scappucciGermaniumQuantumInformation2020}. These include low effective mass which relaxes fabrication constraints \cite{lodariLightEffectiveHole2019}, and strong spin orbit coupling \cite{wangUltrafastCoherentControl2022b}, leading to record-breaking manipulation speeds \cite{liuUltrafastElectricallyTunable2023,froningUltrafastHoleSpin2021b}. Notably, substantial progress has been made in developing hole spin qubits based on strained germanium quantum wells, highlighting the demonstration of high fidelity single- and two-qubit gate \cite{hendrickxFastTwoqubitLogic2020,lawrieSimultaneousSinglequbitDriving2023,zhangUniversalControlFour2025a,tsoukalasDressedSinglettripletQubit2026a}, long coherence time up to 17.6 $\upmu$s \cite{hendrickxSweetspotOperationGermanium2024}, fast qubit control surpassing 100 MHz \cite{hendrickxFastTwoqubitLogic2020} and the implementation of multi-qubit arrays \cite{borsoiSharedControl162024,johnRobustLocalisedControl2025}. 

However, the strong spin-orbit interaction for hole qubits presents a double-edged sword that achieves ultrafast qubit control without the need for auxiliary components like microstrips or micromagnets while exposes qubits to charge noise \cite{scappucciGermaniumQuantumInformation2020,wangOptimalOperationPoints2021}. In addition, the anisotropic spin properties of holes lead to site-dependent and field-dependent coupling to noise sources \cite{piotSingleHoleSpin2022a,mauroGeometryDephasingSweet2024,hendrickxSweetspotOperationGermanium2024,bassiOptimalOperationHole2026,wangOptimalOperationPoints2021}, resulting in both the challenge and opportunity for qubit control. On the one hand, site-dependent $g$ factors enable qubit addressability \cite{hendrickxFastTwoqubitLogic2020} while simultaneously increase the demand for wafer uniformity \cite{vanriggelen-doelmanCoherentSpinQubit2024}. On the other hand, anisotropy offers the possibility of finding a sweet spot that maintains a balance between qubit control and coherence \cite{hendrickxSweetspotOperationGermanium2024,bassiOptimalOperationHole2026}. Recent theoretical works and experimental results have shown that the existence of optimal operation points through choosing the direction and strength of the magnetic field properly or tuning the electrostatic confinement carefully \cite{hendrickxSweetspotOperationGermanium2024,seidlerSpatialUniformityGtensor2025,bassiOptimalOperationHole2026}. These optimal operation points, also known as sweet spots, where hole spin qubits exhibit first-order insensitivity to charge noise \cite{piotSingleHoleSpin2022a}, can effectively extend qubit coherence times. Although sweet spots with enhanced coherence times have been reported previously\cite{hendrickxSweetspotOperationGermanium2024,piotSingleHoleSpin2022a}, a broader experimental understanding of how spin-electric-field susceptibility correlates with qubit coherence and control performance across different magnetic-field orientations is still needed.

Here, we report hole spin qubits in a planar germanium double quantum dots. We experimentally demonstrate the hole $g$ factors of a two-qubit system and quantify their qubit frequency sensitivity to electric fields (longitudinal spin-electric susceptibility, LSES). By varying the orientation of the external magnetic field, we investigate the interplay between charge-noise–induced decoherence, the Rabi frequency $f_{\rm Rabi}$, control fidelities, and magnetic field direction. We find that the longest coherence time ($T_2^{\rm Echo}$) occurs when the qubit frequency exhibits minimal susceptibility to electric fields. This operating point constitutes a local sweet spot for qubit manipulation and results in a fivefold enhancement of $T_2^{\rm Echo}$. The magnetic-field-orientation dependence of the Hahn-echo coherence time $T_2^{\rm Echo}$ closely tracks that of the LSES, providing compelling evidence that qubit decoherence is limited by charge-noise–induced fluctuations of the qubit frequency. Furthermore, at the optimal operating point the qubit is less affected by charge noise. Gate set tomography demonstrates a nearly tenfold suppression of control infidelity, yielding a maximum gate fidelity of 99.82 $\%$, exceeding the threshold required for fault-tolerant quantum computation \cite{fowlerHighthresholdUniversalQuantum2009b}. These results highlight the advantages of hole spin qubits in planar germanium and underscore their strong potential for scalable quantum information processing.

\section{\label{sec:level2}Device fabrication and \MakeLowercase{g} factor anisotropy}

\begin{figure}[b]
\includegraphics{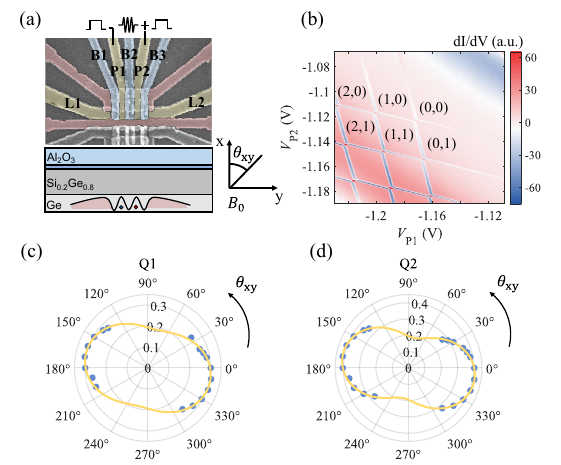}
\caption{\label{fig:1} (a) The lower part of the figure shows false-colored scanning electron microscopy (SEM) images of the double quantum dot (DQD) device. The upper section corresponds to the DQD structure defined by a set of electrostatic gates, while the lower section of the device hosts a single-hole transistor that is used as a charge sensor. Along the white dashed line, a cross-sectional schematic of the device is illustrated in the upper part of the figure. The DQD is electrostatically confined by aluminum gates, including lead gates L1 and L2, barrier gates B1–B3, and plunger gates P1 and P2. Three-stage voltage pulses together with microwave signals are applied to gates P1 and P2 for qubit initialization, manipulation, and readout. The magnetic-field direction parallel to the axis connecting the two quantum dots defines the $x$ direction, while the direction perpendicular to the DQD axis lies in the transverse in-plane direction. The in-plane magnetic-field angle $\theta _{\rm xy}$ is defined with respect to the $x$ axis and is varied starting from this direction. (b) Charge-stability diagram of the DQD. (N$_{\rm 1}$,N$_{\rm 2}$) denote the charge occupations of the quantum dots defined by plunger gates P1 and P2, respectively. The interdot (1,1)-(2,0) charge transition is used as the readout region. Qubit readout is performed using an enhanced latching readout scheme, as also implemented in Ref. \cite{zhouHighfidelityGeometricQuantum2025a}. (c–d) Magnetic-field-orientation dependence of the effective in-plane $g$ factors for the two qubits. The principal axes of the in-plane $g$ tensors for both qubits are tilted by a similar angle of approximately 14$^\circ$ and 12$^\circ$. Despite this similarity, the absolute $g$ factor values differ between the two qubits. For qubit Q1, the $g$ factor varies between 0.19 and 0.30, while for qubit Q2 it ranges from 0.19 to 0.41. These differences indicate distinct confinement potentials and locally varying strain environments for the two single-hole qubits.}
 \end{figure}

Fig.~\ref{fig:1}(a) shows a false-colored scanning electron microscope (SEM) image of the double quantum dots (DQD). The device is fabricated on a high-mobility undoped SiGe/Ge heterostructure, where the electrostatic potential is defined by three overlapping aluminum gate layers, and the ohmic contacts are formed by ion implantation \cite{kongUndopedStrainedGe2023,maSinglespinqubitGeometricGate2024,zhouHighfidelityGeometricQuantum2025a}. The fabrication process is similar to that reported in Refs.~\cite{zhangGiantAnisotropySpin2020c,huFloppingmodeSpinQubit2023c}. The measurements are performed in a dilution refrigerator with a base temperature of 15 mK, and details of the instrumentation are provided in Appendix~\ref{App-A}.

The device is divided into two functional sections. A DQD is defined beneath gates P1 and P2, while gates B1–B3 are used to tune both the interdot tunnel coupling and the tunnel rates between the quantum dots and their reservoirs. A separate quantum dot on the opposite side of the DQD serves as a charge sensor. An external in-plane magnetic field $B_{0}=900$ mT is applied to induce Zeeman splitting and defines the spin qubits \cite{niSWAPGateSpin2025}. Microwave pulses and three-step voltage pulse sequences are applied to gates P1 and P2 to initialize, manipulate, and read out the qubit states.

By sweeping the voltages on gates P1 and P2, we obtain the charge stability diagram shown in Fig.~\ref{fig:1}(b). Regions with different charge occupations $(N_{1},N_{2})$ are clearly resolved using the charge sensor, where $N_{i}$ denotes the number of holes in the quantum dot under gate Pi. The absence of tunneling lines in the upper-right corner indicates full depletion of the DQD. We operate the device in the $(1,1)$–$(2,0)$ charge transition regime and employ latched Pauli spin blockade for spin-state readout. The readout mechanism and signal are similar to those reported in Ref.~\cite{zhouHighfidelityGeometricQuantum2025a}.

For holes confined in quasi-two-dimensional systems, such as a strained germanium quantum well, size quantization lifts the fourfold degeneracy of the $j=3/2$ valence band, leading to a splitting between heavy-hole (HH) and light-hole (LH) states \cite{stanoQuantificationHeavyholeLighthole2025,scappucciGermaniumQuantumInformation2020}. The degree of HH–LH mixing is governed by the strength of the electrical confinement both along the growth direction and within the plane of the quantum well, which in turn gives rise to a pronounced anisotropy of the hole $g$ factor \cite{jirovecDynamicsHoleSingletTriplet2022,saez-mollejoExchangeAnisotropiesMicrowavedriven2025}. To probe this anisotropy, we measure the hole spin resonance frequencies of two qubits, denoted as $f_{\rm Q1}$ and $f_{\rm Q2}$, while rotating the external magnetic field $B_0$ within the $xy$ plane. From the angular dependence of the resonance frequencies, the effective $g$ factors of the two qubits, $g_{\rm Q1}$ and $g_{\rm Q2}$, are extracted using $g_{\rm Qi} = hf_{\rm{Qi}}/\mu_{B}B_0$, where $h$ is Planck’s constant and $\mu_{B}$ is the Bohr magneton. 

 As shown in Fig.\ref{fig:1}(c-d), the resulting in-plane $g$ factor maps reveal a strong anisotropy of the Zeeman splitting. In Fig.\ref{fig:1}(c-d), for qubit Q1 (Q2), the $g$ factor reaches a maximum value of $g_{\rm Q1}=0.30 \pm 1.6\times10^{-5}$ ($g_{\rm Q2}=0.41 \pm 2.3\times10^{-5}$) when the magnetic field is aligned close to the $x$ axis, which is perpendicular to the interdot axis, and a minimum value of $g_{\rm Q1}=0.19$ ($g_{\rm Q2}=0.19$) when the field is aligned close to the $y$ axis, parallel to the interdot axis.

According to the numerical simulation (yellow solid lines shown in Fig.\ref{fig:1}(c-d)) \cite{sarkarElectricalOperationPlanar2023,piotSingleHoleSpin2022a}, we evaluate that the origin of $g$ factor anisotropy comes from the asymmetric potential in x and y direction and non-circular confinement of quantum dots\cite{marieHoleSpinQuantum1999,sarkarElectricalOperationPlanar2023,jirovecDynamicsHoleSingletTriplet2022,trifonovHomogeneousOpticalAnisotropy2021}. In Fig.~\ref{fig:1}(c–d), by comparing the maximum and minimum $g$ factors of Q1 and Q2, we observe that the differences in the oscillation amplitudes of the $g$ factor in the DQD arise from unequal squeezing induced by the barrier and confinement gates under enhanced latching readout conditions. In addition, the in-plane maps suggest that the principal axes of $g$ factor of two qubits is similar but is slightly misaligned with the magnet axis (about $14 ^{\circ}$ for Q1 and $12 ^{\circ}$ for Q2 ), probably resulting from the strain generated during the device fabrication process\cite{abadillo-urielHoleSpinDrivingStrainInduced2023,wangOptimalOperationPoints2021,seidlerSpatialUniformityGtensor2025a}. 

\begin{figure}[t]
\includegraphics{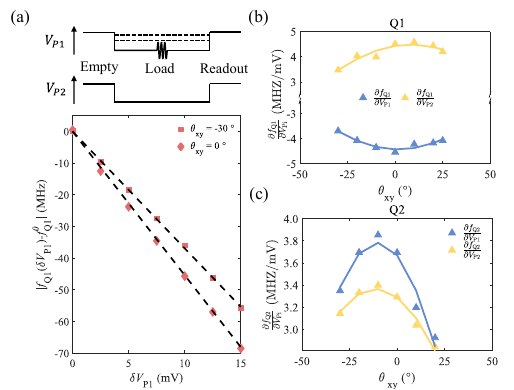}
\caption{\label{fig:2} (a) Top: Pulse sequence used to measure the longitudinal spin–electric susceptibility, LSES$_{\rm Q1}$ and LSES$_{\rm Q2}$. The sequence consists of three stages—Empty, Load, and Readout—corresponding to the charge configurations (1,0), (1,1) and (1,0), respectively. Qubit manipulation is performed during the Load stage. By varying the voltage applied to plunger gate P1 or P2, the resonance frequency of qubit Q1 and Q2 is shifted accordingly. The resulting gate-voltage-dependent qubit frequency shifts are extracted and plotted, as shown in the lower panel of (a). The frequency shifts exhibit a linear dependence on the applied gate voltage, from which the LSES is obtained. The lower panel shows representative examples of LSES$_{\rm P1}$ measured at magnetic-field orientations $\theta _{\rm xy} = -30^{\circ}$ and $\theta _{\rm xy} = 0^{\circ}$. (b-c) Angular dependence of LSES for qubits Q1 and Q2 as a function of the in-plane magnetic-field angle $\theta_{\rm xy}$. For qubit Q1, LSES exhibits both positive and negative values for gates P1 (blue triangles) and P2 (yellow triangles), indicating opposite influences of the two gates on Q1. Both $\frac{\partial f_{\rm Q1}}{\partial V_{\rm P1}}$ and $\frac{\partial f_{\rm Q1}}{\partial V_{\rm P2}}$ show a pronounced parabolic dependence on $\theta_{\rm xy}$, with a maximum at $\theta_{\rm xy}=0^{\circ}$. As the magnetic-field orientation deviates from this angle, LSES$_{\rm Q1}$ decreases, reflecting a reduced spin–electric susceptibility and suppressed charge-noise-induced dephasing. A similar angular dependence is observed for qubit Q2. Notably, $\frac{\partial f{\rm Q2}}{\partial V_{\rm P1}}$ (blue triangles) exceeds $\frac{\partial f_{\rm Q2}}{\partial V_{\rm P2}}$ (yellow triangles). This behavior is attributed to the asymmetric tunnel rates between the two individual quantum dots and the source and drain, which strongly squeeze the quantum dot under gate P1 and modify the effective confinement, highlighting a strong gate-geometry-dependent response. LSES$_{\rm Q2}$ remains strictly positive over the entire angular range, implying that voltages applied to both P1 and P2 enhance the in-plane $g$ factor of Q2. The maximum of LSES$_{\rm Q2}$ occurs at $\theta_{\rm xy}=-10^{\circ}$.}
\end{figure}

\section{\label{sec:level3}spin-electric susceptibility} 

Considering that the anisotropy of $g$ factor originates from HH-LH mixing, which is related to variation in the electrical potential \cite{stanoQuantificationHeavyholeLighthole2025,jirovecDynamicsHoleSingletTriplet2022}. It is natural to expect that the qubit Larmor frequency is likewise modulated by electric fields. This effect can be quantitatively characterized by the longitudinal spin–electric susceptibility (LSES), which serves as a direct measure of the sensitivity of spin qubits to charge noise\cite{piotSingleHoleSpin2022a,hendrickxSweetspotOperationGermanium2024}. Therefore, we measure the change in Larmor frequency as a function of the voltage applied to the plunger gate P1 and P2. we define the LSES as $\rm{LSES}_{\rm{Q1}} =\it\frac{\partial f{\rm Q1}}{\partial V_{\rm Pi}}$ and $\rm{LSES}_{\rm{Q2}} = \it \frac{ \partial f_{\rm{Q2} }}{ \partial V_{\rm{Pi}}}$, where i = 1,2.

As an example, we describe the extraction of $\frac{ \partial f_{\rm{Q1} }}{ \partial V_{\rm{P1}}}$. As shown in the upper panel of Fig.~\ref{fig:2}(a), a single-shot measurement cycle is divided into three stages: empty, load, and read. To probe the response of $f_{\rm{Q1}}$ to gate P1, we vary the pulse voltage applied to P1 during the load stage and measure the corresponding spin resonance frequency $f_{\rm{Q1}}$. The resulting shift in $f_{\rm{Q1}}$ is plotted as a function of $\delta V_{\rm{P1}}$ in the lower panel of Fig.~\ref{fig:2}(a). To illustrate the LSES more intuitively, measurements are performed for two representative in-plane magnetic field orientations $\theta_{\rm{xy}} = -30^{\circ}, 0^{\circ}$. As expected, the change in $f_{\rm{Q1}}$ exhibits an approximately linear dependence on $\delta V_{\rm{P1}}$. The red squares and diamonds denote the experimental data, while the dashed lines represent linear fits, from which the slope $\frac{ \partial f_{\rm{Q1}} }{ \partial V_{\rm{P1}}}$ is extracted. Applying the same procedure to obtain $\frac{ \partial f_{\rm{Q1}} }{ \partial V_{\rm{P2}}}$, as well as the corresponding LSES$_{\rm Q2}$, we determine the $\theta_{\rm{xy}}$ dependence of the LSES for both Q1 and Q2, as shown in Fig.~\ref{fig:2}(b)–(c).

Noticeably, the value of $\rm{LSES}_{\rm{Q1}}$ can be positive or negative as shown in Fig.\ref{fig:2}(b). When increasing $V_{\rm{P1}}$, we speculate the hole wave function extends proportionally because P1 is the gate directly above Q1, resulting in the decrease of $g$ factor. In contrast to increasing $V_{\rm{P2}}$, owing to the position of P2 relative to Q1 on the side and relative to Q2 in the vertical position, the expansion of the hole wave function of Q2 is equivalent to the compression of the wave function of Q1, which will lead to the increase of $g_{x}$ and $g_{y}$. In the event that we consider the absolute value of $\rm{LSES}_{\rm{Q1}}$, the variation trend of $\rm{LSES}_{\rm{Q1}}$ near the $\theta_{\rm{xy}}$ of 0$^{\circ}$ exhibits a parabolic shape. When the orientation of magnetic field points from $x$ axis to $+$ $y$ axis or $-$ $y$ axis, the $\rm{LSES}_{\rm{Q1}}$ decreases, which means the most sensitive position approaches $\theta_{\rm{xy}} = 0^{\circ}$. 

In contrast, $\rm{LSES}_{\rm{Q2}}$ remains positive over the entire angular range, as shown in Fig.~\ref{fig:2}(c). We further observe that $\frac{ \partial f_{\rm{Q2}} }{ \partial V_{\rm{P1}}}$ is consistently larger than $\frac{ \partial f_{\rm{Q2}} }{ \partial V_{\rm{P2}}}$, despite the fact that P2 is the plunger gate directly associated with Q2. This behavior can be attributed to the asymmetric tunnel coupling between Q1 and Q2 introduced by the enhanced latching readout configuration. As a result, the hole wave function in Q2 is strongly squeezed in the $x$–$y$ plane. At the same time, the vertical confinement in the planar germanium quantum well is inherently strong. Consequently, increasing $V_{\rm{P2}}$ does not substantially modify the wave function of Q2 along the growth direction, and the tunability of the electrostatic potential via P2 is therefore limited. In contrast, gate P1 is located laterally with respect to Q2, such that variations in $V_{\rm{P1}}$ more efficiently perturb the in-plane confinement potential of Q2, leading to a larger modulation of the qubit Larmor frequency. Similar to Q1, the angular dependence of $\rm{LSES}_{\rm{Q2}}$ exhibits a parabolic behavior at $\theta{\rm{xy}} \approx -10^{\circ}$. When the magnetic field is rotated away from this angle toward either more negative or more positive orientations, the magnitude of $\rm{LSES}_{\rm{Q2}}$ decreases, indicating a reduced sensitivity of the qubit frequency to electric-field fluctuations.

\begin{figure}[b]
\includegraphics{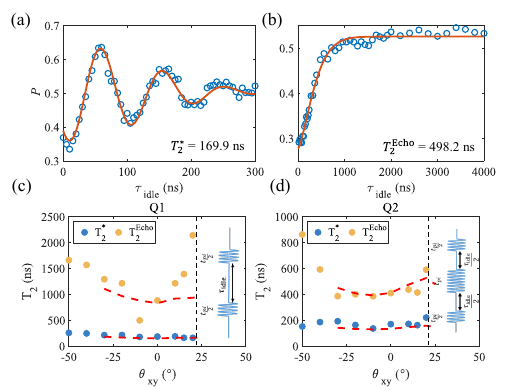}
\caption{\label{fig:3} (a) Ramsey fringe measured as a function of the idle time $\tau_{\rm idle}$ with a detuning of 10 MHz, yielding a fitted dephasing time $T_2^* = 169.9,\mathrm{ns}$ at $\theta_{\rm xy}=0^{\circ}$ for qubit Q1. (b) Hahn-echo fringe measured by varying $\tau_{\rm idle}$, with a fitted decoherence time $T_2^{\rm Echo} = 498.2,\mathrm{ns}$ at $\theta_{\rm xy}=0^{\circ}$ for qubit Q1. In both panels (a-b), the orange solid lines denote the fits to the experimental data.
(c) Dephasing time $T_2^*$ and Hahn-echo coherence time $T_2^{\rm Echo}$ of qubit Q1 as functions of the in-plane magnetic-field angle $\theta_{\rm xy}$. While $T_2^*$ remains essentially constant over the measured angular range, $T_2^{\rm Echo}$ shows a non-monotonic dependence, first decreasing and then increasing with $\theta_{\rm xy}$. This trend closely follows the angular dependence of LSES$_{\rm Q1}$ in Fig.~\ref{fig:2}(b), with the longest (shortest) $T_2^{\rm Echo}$ occurring at the minimum (maximum) LSES${\rm Q1}$. The maximum $T_2^{\rm Echo}$ is approximately five times larger than the minimum value. The simulations for $T_2^*$ and $T_2^{\rm Echo}$ for Q1 are shown as red dashed lines (see Appendix~\ref{App-B} for details).
(d) $T_2^*$ and $T_2^{\rm Echo}$ of Q2 as functions of $\theta_{\rm xy}$. The angular dependence of $T_2^*$ is similar to that of Q1. In contrast, $T_2^{\rm Echo}$ remains nearly constant for $-30^{\circ} \le \theta_{\rm xy} \le 10^{\circ}$ and increases outside this range. The longest $T_2^{\rm Echo}$ is observed at the magnetic-field orientation corresponding to the minimum LSES${_{\rm Q2}}$ which characterized in Fig.~\ref{fig:2}(c). The red dashed lines represent the simulated coherence times $T_2^{*}$ and $T_2^{\rm Echo}$ for qubit Q1, respectively (see Appendix~\ref{App-B} for simulation details).
} 
\end{figure}

\section{\label{sec:level4}Magnetic Field-Dependent Properties of Qubits}

Previous studies have shown that qubit decoherence is predominantly limited by fluctuations of the Larmor frequency induced by charge noise\cite{noiriFastUniversalQuantum2022a,philipsUniversalControlSixqubit2022e,yonedaNoisecorrelationSpectrumPair2023}. To investigate the relationship between the LSES and qubit coherence time, We focus on how the coherence time of hole spins depends on the orientation of the magnetic field ($\theta_{\rm{xy}}$). We note that varying $\theta_{\rm{xy}}$ modifies the effective $g$ factor and thus shifts the qubit resonance frequency, which in turn affects the microwave response at different frequencies. To exclude this effect in the following measurements, we keep the qubit resonance frequency fixed while varying $\theta_{\rm{xy}}$. 

We first perform Ramsey measurements with a detuning of 10 MHz from the qubit resonance to characterize the dephasing time $T_{2}^{*}$ for both qubits. As shown in inset of Fig.~\ref{fig:3}(c), the Ramsey sequence consists of two $\pi/2$ pulses separated by a variable idle time $\tau_{\rm{idle}}$. Fig.~\ref{fig:3}(a) shows a representative Ramsey decay for Q1, where the singlet probability $P$ is plotted as a function of $\tau_{\rm{idle}}$. By fitting the data with $A*\rm{exp}(-(\tau_{\rm{idle}}/T_2^{*})^2)*\sin(2\pi f(\tau_{\rm{idle}}-\tau_{0}))+B$. we extract $T_{2}^{*} = 169.9~\text{ns}$ at $\theta_{\rm{xy}} = 0^{\circ}$. Remarkably, the extracted $T_{2}^{*}$ values for both Q1 and Q2 remain essentially constant over the full range of $\theta_{\rm{xy}}$, as shown in Fig.~\ref{fig:3}(c)–(d) (blue solid circles). This observation indicates that the magnetic-field–orientation dependence of $T_2^*$ is effectively isotropic. We attribute this behavior to the dominance of low-frequency noise during the long acquisition times and extensive signal averaging required for Ramsey measurements, which can obscure any weak anisotropy in $T_{2}^{*}$ \cite{piotSingleHoleSpin2022a,noiriFastUniversalQuantum2022a}.

In contrast to the Ramsey experiment, we employ the Hahn-echo technique to suppress the effects of low-frequency noise\cite{kawakamiElectricalControlLonglived2014b}. The Hahn-echo sequence incorporates a $\pi$ pulse inserted halfway through the total waiting time $\tau_{\rm{idle}}$, thereby dividing it into two equal segments, as shown in inset of Fig.~\ref{fig:3}(d). This refocusing pulse effectively cancels qubit dephasing arising from quasi-static and low-frequency fluctuations\cite{hendrickxSweetspotOperationGermanium2024,cywinskiHowEnhanceDephasing2008b,madzikControllableFreezingNuclear2020b}. The decay of the echo amplitude is fitted with a stretched exponential function, $\rm{exp}(-(\tau_{\rm{idle}}/T_2^{\rm Echo})^{\beta})$, where $T_2^{\rm Echo}$ denotes the Hahn-echo coherence time and the exponent $\beta$ is treated as a free fitting parameter. Fig.~\ref{fig:3}(b) shows a representative Hahn-echo decay for Q1, from which we extract $T_2^{\rm Echo} = 498.2~\mathrm{ns}$ and the corresponding value of $\beta$. The fitted exponent $\beta$ provides insight into the spectral properties of the dominant noise source. In particular, the noise can be characterized by a power spectral density of the form $S(f) = S_{\rm{hf}}(f_0/f)^{\alpha}$ with the exponent $\alpha = \beta - 1$.

Next, to investigate the relationship between the magnetic-field orientation and the Hahn-echo coherence time $T_2^{\rm Echo}$, we measure $T_2^{\rm Echo}$ for both qubits as a function of $\theta_{\rm{xy}}$, as shown in Fig.~\ref{fig:3}(c)–(d) (yellow solid circles). For both Q1 and Q2, we observe a pronounced anisotropy in $T_2^{\rm Echo}$ with respect to the magnetic-field orientation. Specifically, $T_2^{\rm Echo}$ for Q1 (Q2) increases from approximately 500 ns (400 ns) to 2200 ns (800 ns), corresponding to an enhancement by a factor of about five (two). Notably, the minimum coherence time coincides with the maximum of $\rm{LSES}_{\rm{Q1}}$ and $\rm{LSES}_{\rm{Q2}}$. This strong correlation provides compelling evidence that qubit decoherence is predominantly limited by charge-noise–induced fluctuations of the qubit resonance frequency.

The qubit quality factor $Q$ is a key metric for qubit manipulation, as it captures the balance between coherence and control speed. Therefore, it is essential to characterize both the Rabi frequency $f_{\rm Rabi}$ and the quality factor $Q$ (see Appendix~\ref{App-C} for details). We find that for qubit Q2, the echo-based quality factor $Q^{E}$, defined as $Q^{E}=2 T_2^{\rm Echo} f_{\rm Rabi}$, reaches its maximum at the optimal operating point, where it is approximately five times larger than in the least favorable configuration. In addition, $f_{\rm Rabi}$ for Q1 and Q2 exhibits markedly different dependencies on the magnetic-field orientation. This behavior is likely influenced by the distinct confinement potentials of the two quantum dots, together with the fact that the microwave drive is applied only to gate P2. Consequently, the two qubits experience different microwave coupling strengths and may be governed by different control mechanisms \cite{rodriguez-menaLinearinmomentumSpinOrbit2023}.

To quantitatively demonstrate the magnetic-field–orientation dependence of the dephasing time $T_2^*$ and the Hahn-echo coherence time $T_2^{\rm Echo}$, we model the decoherence induced by charge noise under the assumption that voltage fluctuations on different gates are uncorrelated \cite{piotSingleHoleSpin2022a}. In this analysis, we focus on the dominant contributions from plunger gates P1 and P2, and assume that the total charge noise experienced by the qubit arises from the sum of the noise contributions associated with these two gates (More details are shown in Appendix \ref{App-B}). By substituting the experimentally extracted longitudinal spin–electric susceptibilities from Fig.~\ref{fig:2}(b–c) into the expression in Appendix \ref{App-B} and treating the high-frequency noise amplitudes $S_{\rm{Pi}}^{\rm{hf}}$ and $S_{\rm{Pi}}^{\rm{lf}}$ as fitting parameters, we calculate the theoretical Hahn-echo decoherence rates for both qubits. 

Through simulations of qubits Q1 and Q2, we find that the simulated dephasing times $T_2^*$ are in good agreement with the experimental results, indicating that low-frequency charge noise is an important contributor to $T_2^*$ \cite{piotSingleHoleSpin2022a}. In contrast, the numerically calculated coherence times deviate from the experimentally measured $T_2^{\rm Echo}$ (see Appendix~\ref{App-B} for more details). This discrepancy can be attributed to voltage fluctuations on the barrier gates, which are expected to contribute to both dephasing and decoherence. The LSES of qubit affected by barrier gate voltage may much larger and significant in simulation \cite{hendrickxSweetspotOperationGermanium2024}. 
However, fast voltage pulses can only be applied to the plunger gates P1 and P2, preventing experimental access to the longitudinal spin–electric susceptibilities associated with the barrier gates B1–B3. Consequently, noise contributions from these gates are not included in the present simulation, which may contribute to the residual mismatch between the numerical model and the experimental data.

\begin{figure*}[t]
\centering
\includegraphics{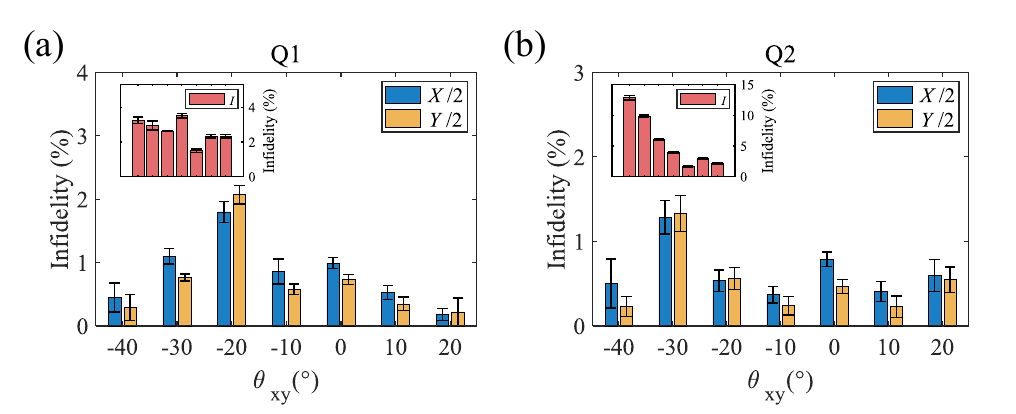}
\caption{\label{fig:4} (a–b) Control infidelities of the gate set ${I/2, X/2, Y/2}$ as functions of the in-plane magnetic-field angle $\theta_{\rm xy}$. The infidelities of the $X/2$ and $Y/2$ gates exhibit non-monotonic angular dependence, first increasing and then decreasing with $\theta_{\rm xy}$. The maximum gate error occur at magnetic-field orientations corresponding to the shortest $T_2^{\rm Echo}$. For qubit Q1 (Q2), the control infidelity at the optimal operating point is reduced by approximately one order of magnitude (a factor of five) compared to the worst case. The insets show the $I$ gate infidelity. The horizontal axis is $\theta_{\rm{xy}}$, and the vertical axis is the gate infidelity, as in the main panel. Although the $I$ gate infidelity is determined by both $T_2^*$ and $f_{\rm Rabi}$, the nearly constant $T_2^*$ across the measured range of $\theta_{\rm{xy}}$ indicates that the observed angular dependence is governed predominantly by $f_{\rm Rabi}$. For Q1, the $I$-gate infidelities exhibits a non-monotonic dependence on $\theta_{\rm xy}$, consistent with the behavior of $f_{\rm Rabi}$ in Appendix~\ref{App-C}. For Q2, the infidelities initially decreases and then saturates, following the trend of $f_{\rm Rabi}$ in Appendix~\ref{App-C}. This saturation indicates that further enhancement of $f_{\rm Rabi}$ yields diminishing returns for gate fidelity, and that additional improvements require optimization of other qubit coherence properties, such as $T_2^*$ and $T_2^{\rm Rabi}$.}
\end{figure*}

\section{\label{sec:level5}Magnetic field-dependent qubit Control fidelity }

Building on the systematic characterization of the anisotropic coherence properties of the qubits, we next evaluate the control fidelities of qubits Q1 and Q2 using gate set tomography (GST) \cite{nielsenGateSetTomography2021a,blume-kohoutDemonstrationQubitOperations2017}. Compared with conventional quantum process tomography (QPT), GST mitigates the impact of state preparation and measurement (SPAM) errors, enabling a more reliable extraction of gate fidelities. Unlike randomized benchmarking (RB) \cite{chowRandomizedBenchmarkingProcess2009}, GST explicitly models state preparation and measurement and, through maximum-likelihood estimation, reconstructs the complete error structure of the gate set \cite{nielsenGateSetTomography2021a}. This capability allows simultaneous identification of state preparation and measurement errors, Hamiltonian miscalibrations, and stochastic noise processes, making GST both a benchmarking and diagnostic tool for optimizing qubit control \cite{xueQuantumLogicSpin2022a,huangHighfidelitySpinQubit2024,tanttuAssessmentErrorsHighfidelity2024}.

We systematically characterize the gate-set control infidelities of qubits Q1 and Q2 for in-plane magnetic field orientations within a range of $\pm 50^{\circ}$ around $\theta_{\rm{xy}} = 0^{\circ}$. The gate set employed in the GST measurements is ${X/2, Y/2, I}$. Here, the $X/2$ ($Y/2$) gate corresponds to a rotation by an angle $\pi/2$ about the $X$ ($Y$) axis with a gate duration of $t_{\pi/2}$. The $I$ gate denotes the identity operation and is experimentally implemented by waiting for the same duration $t_{\pi/2}$.

Fig.~\ref{fig:4}(a) and (b) show the gate-set control infidelities of qubits Q1 and Q2, respectively, extracted using the GST protocol. For qubit Q1, as shown in Fig.~\ref{fig:4}(a), the control infidelities of the $X/2$ and $Y/2$ gates exhibit a pronounced maximum around $\theta_{\rm{xy}} \approx -20^{\circ}$ as the in-plane magnetic field orientation is varied. When the magnetic field angle deviates from this position in either direction, the gate infidelities decrease. Within the measured angular range, at the optimal operating point, the gate infidelity is reduced by approximately one order of magnitude, while the maximum gate fidelity reaches 99.82$\%$, exceeding the threshold required for fault-tolerant quantum computation.
Notably, the angular position corresponding to the minimum gate infidelities coincides with and the minimum LSES for qubit Q1. Moreover, the overall angular dependence of the gate infidelities closely follows that of $T_2^{\rm Echo}$, indicating a direct connection between qubit coherence and high-fidelity control.

For qubit Q2, the $X/2$ and $Y/2$ gate infidelities remain nearly constant over most magnetic-field orientations, with only a pronounced maximum appearing around $\theta_{\rm{xy}} \approx -30^{\circ}$. These angles corresponds to the regime where the Hahn-echo coherence time $T_2^{\rm Echo}$ begins to increase as a function of the magnetic-field orientation. In the range $\theta_{\rm{xy}} \in (-30^{\circ}, 10^{\circ}]$, the coherence time $T_2^{\rm Echo}$ of Q2 remains essentially unchanged, and correspondingly the gate infidelities exhibit little angular dependence. In contrast, for $\theta_{\rm{xy}} \in (-50^{\circ}, -30^{\circ}]$, the coherence time becomes longer while the gate infidelity decreases, with the infidelity at the optimal operating point reduced by approximately a factor of five.

In contrast to the behavior of the $X$/2 and $Y$/2 gates, the infidelities of the $I$ gate for both qubits do not exhibit an angular dependence similar to that of the Hahn-echo coherence time $T_2^{\rm Echo}$. This difference arises because the $I$ gate, implemented as an idle operation with a duration of $t_{\pi/2}$, is primarily limited by the interplay between the inhomogeneous dephasing time $T_2^{*}$ and the Rabi frequency $f_{\rm Rabi}$\cite{noiriFastUniversalQuantum2022a}. As the in-plane magnetic field orientation $\theta_{\rm{xy}}$ is varied from $-50^{\circ}$ to $50^{\circ}$, the extracted $T_2^{*}$ values for both qubits remain nearly constant. The Rabi frequency of Q1 exhibits a non-monotonic dependence on $\theta_{\rm{xy}}$, whereas that of Q2 increases monotonically (Appendix.~\ref{App-C}). Consequently, the $I$-gate infidelities follow the corresponding trends of $f_{\rm Rabi}$. For Q1, the $I$-gate infidelities first increases and then decreases, while for Q2 it decreases gradually and eventually saturates. Once $f_{\rm Rabi}$ becomes sufficiently large, the gate performance is no longer limited by the driving strength but instead by low-frequency noise, which cannot be further suppressed by increasing $f_{\rm Rabi}$ \cite{noiriFastUniversalQuantum2022c}. This mechanism leads to a saturation of the $I$-gate fidelity, a behavior that is particularly pronounced for qubit Q2.

\section{\label{sec:level6}CONCLUSION}

In this work, we have investigated the anisotropic properties of planar germanium hole spin qubits arising from the orientation of the external magnetic field. We demonstrated that the enhancement of coherence time and the reduction of control infidelity exhibit pronounced and correlated anisotropies as the magnetic-field direction is varied within the planar germanium.

The strong intrinsic spin–orbit coupling of valence-band holes in planar germanium enables fast, all-electrical qubit manipulation without the need for micromagnets or microwave antennas, making this platform particularly attractive for scalable quantum computing architectures. At the same time, spin–orbit coupling renders the qubit susceptible to charge noise. Owing to the highly anisotropic nature of both the $g$ tensor and the spin–orbit interaction, the spin–electric coupling itself becomes strongly orientation dependent, opening the possibility to optimize qubit performance through magnetic-field engineering.

By quantitatively extracting the spin–electric susceptibility from the gate-voltage dependence of the qubit resonance frequency, we identify a local magnetic-field orientation at which the susceptibility is minimized and the coherence time is maximized. At this optimal field direction, the Hahn-echo coherence time is enhanced by approximately a factor of five. Analysis of the coherence decay yields a noise spectral exponent close to $1/f^{0.907}$, consistent with the noise spectrum characterized by Ramsey and CPMG experiments (see Appendix~\ref{App-B} for details). By fitting the angular dependence of the coherence time, we further disentangle that decoherence is dominated by low-frequency noise.

We further show that qubit control infidelities exhibit the same anisotropic trends. The infidelities of $X/2$ and $Y/2$ gates are primarily limited by charge-noise–induced decoherence and therefore closely follow the angular dependence of the Hahn-echo coherence time $T_2^{\rm Echo}$. In contrast, the infidelities of the idle $I$ gate initially decreases and then saturates, reflecting the interplay between the dephasing time $T_2^{*}$ and the Rabi frequency $f_{\rm Rabi}$. At the optimal operating point, the control infidelities of the single-qubit gates for Q1 and Q2 are reduced by approximately one order of magnitude and a factor of five compared to the worst-case direction, respectively. The maximum control fidelities reach 99.82 $\%$ for the $X/2$ gate and 99.78 $\%$ for the $Y/2$ gate, exceeding the threshold required for fault-tolerant quantum computation.

Our experiments systematically characterize the anisotropic $g$ factor, coherence time, quality factor, and gate infidelities of planar germanium hole spin qubits, and correlate their variations with the longitudinal spin-electric susceptibility. These results demonstrate that optimizing the magnetic-field orientation can significantly prolong the coherence time while substantially reducing the gate infidelity, highlighting the importance of magnetic-field engineering for controlling qubit performance.

For future large-scale qubit arrays, magnetic-field engineering, including control over the field strength, direction, spatial profile, and location \cite{4lky-413f,RN54,RN56}, may provide a promising avenue for scalable qubit control. More broadly, engineering the spin–orbit field through complementary approaches may offer additional flexibility in optimizing qubit performance. Theoretical studies have shown that strain engineering \cite{PhysRevLett.131.097002,PhysRevB.108.205416}, optimization of the crystal growth direction \cite{PhysRevB.103.085309}, engineering of the strength and spatial distribution of the electric field \cite{PhysRevB.106.235426}, and optimization of the quantum-dot structure and geometry \cite{secchi2026holespinqubitsgermaniumsingleparticle,sarkarElectricalOperationPlanar2023} can enhance the spin–orbit coupling and enable greater tunability of its strength and direction. The combined engineering of the magnetic field, electric field, and strain could therefore provide greater flexibility in controlling LSES and, consequently, in optimizing qubit coherence and manipulation performance in large-scale qubit arrays.

\begin{acknowledgments}
This work was supported by the National Natural Science Foundation of China (Grants No. 12474490, 62404248 and 12574552), Quantum Science and Technology-National Science and Technology Major Project (Grant No. 2021ZD0302300). This work was partially carried out at the USTC Center for Micro and Nanoscale Research and Fabrication.
\end{acknowledgments}

\appendix

\section{\label{App-A}Measurement setup}

All measurements were carried out in an Oxford dry dilution refrigerator operating at a base temperature of approximately 15 mK. Gate-voltage pulse sequences consisting of three stages were produced using a Keysight M8190 arbitrary waveform generator (AWG). The pulsed signals were combined with DC gate voltages via an analog summing amplifier (SRS SIM980) and subsequently applied to the plunger gates through the DC port of a commercial bias tee (Anritsu K251). Microwave excitation was delivered through the RF port of the same bias tee by applying a signal generated from a Keysight E8267D vector signal generator. This microwave signal was I/Q modulated using channel pairs of a Tektronix AWG5208. Pulse modulation was implemented such that the modulation signal was switched on 500 ns prior to the microwave burst and switched off 500 ns after the burst. The qubit state was detected using charge sensing by monitoring the current through the single hole transistor (SHT). The output current signal was amplified at room temperature using low-noise current and voltage amplifiers (SRS SR570 and SR560) and subsequently digitized by a PCI-based waveform digitizer (AlazarTech ATS9440) with a sampling rate of 5 MSa/s.

\section{\label{App-B} The theoretical simulations of $T_2^*$ and $T_2^{\rm Echo}$}

According to the theoretical analysis in Ref.~\cite{piotSingleHoleSpin2022a}, the dephasing experienced by the spin caused by voltage noise leads to a decay of spin coherence, reflected in the suppression of the off-diagonal elements of the density matrix in the rotating frame. The exponent describing the decay of the off-diagonal elements can be expressed as $-\frac{1}{2}\left\langle \delta \phi_{R}(t)^2 \right\rangle$.

We have the functions to describe the effects of the pulse sequence performed over frequency interval. For Ramsey sequence

\begin{equation}
\left|\tilde{\eta}_t^{\mathrm{R}}(f)\right|^2
=
\left(
\frac{\sin(\pi f t)}{\pi f}
\right)^2 .
\end{equation}
By substituting Eq. (B1) into Eq. (B2),

\begin{equation}
\exp\!\left(-\frac{1}{2}\left\langle \delta \phi_{R}(t)^2 \right\rangle\right)
=
\exp\!\left[
-\left(\frac{t}{T_2^{*}}\right)^2
\right] .
\end{equation}
the resulting dephasing rate for Q1 can be expressed as 

\begin{equation}
\frac{1}{T_2^{*}} \approx 2\pi
\sqrt{
\ln\!\left(\frac{f_h}{f_l}\right)
\, f_0
\sum_{i}(\frac{ \partial f_{\rm{Q1} }}{ \partial V_{\rm{Pi}}})^{2}S_{\rm{Pi}}^{\rm{lf}}\,
}\, 
\end{equation}

The voltage noise spectral density on gate Pi is modeled as $S_{\rm{Pi}}(f) = S_{\rm{Pi}}^{\rm{hf}}(f_0/f)^{\alpha}$, with the noise exponent $\alpha = \beta - 1 = 0.907$ as extracted from the Hahn-echo decay. Following the same procedure, for the Hahn echo sequence,

\begin{equation}
\left|\tilde{\eta}_t^{\mathrm{E}}(f)\right|^2
=
\frac{\sin^4\!\left(\pi f t / 2\right)}{\left(\pi f / 2\right)^2} .
\end{equation}
By substituting Eq. (B3) into Eq. (B4)

\begin{equation}
\exp\!\left(-\frac{1}{2}\left\langle \delta \phi_{E}(t)^2 \right\rangle\right)
=
\exp\!\left[
-\left(\frac{t}{T_2^{\rm Echo}}\right)^{\alpha +1}
\right] .
\end{equation}
Taking into account the Hahn-echo filter function, the resulting decoherence rate for Q1 can be expressed as 
\begin{equation}
\frac{1}{T_2^{\rm Echo}} \approx 4.91f_{0}^{1/0.907}(\sum_{i}(\frac{ \partial f_{\rm{Q1} }}{ \partial V_{\rm{Pi}}})^{2}S_{\rm{Pi}}^{\rm{hf}})^{0.53}
\end{equation}

For Q2, the expression for the dephasing rate and decoherence rate has the same functional form as that for Q1, with the only difference being that $\frac{ \partial f_{\rm{Q1}} }{ \partial V_{\rm{Pi}}}$ is replaced by $\frac{ \partial f_{\rm{Q2}} }{ \partial V_{\rm{Pi}}}$.

\begin{figure}[htbp]
\includegraphics{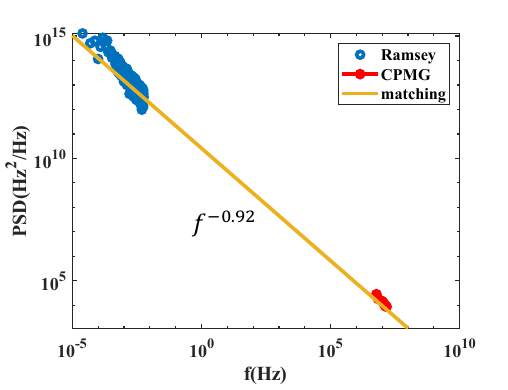}
\caption{\label{fig:5} Noise spectrum extracted from Ramsey measurements over a laboratory timescale of 55 h (blue circles) and from dynamical decoupling experiments using CPMG sequences (red circles). The solid line corresponds to a $1/f^{0.92}$ noise spectrum.
}
\end{figure}

Using Ramsey experiments with a qubit frequency detuning of 10 MHz, we extract the temporal fluctuations of the qubit resonance frequency over a laboratory timescale of 55 h. In addition, by applying dynamical decoupling sequences based on Carr–Purcell–Meiboom–Gill (CPMG) protocols, we measure the coherence time $T_2^{\rm CPMG}$ for different decoupling orders $N$. From these measurements, we reconstruct the noise spectrum over a broad frequency range, as shown in Fig.~\ref{fig:5}. 

From the extracted noise spectrum, we obtain approximate values for the low-frequency and high-frequency noise amplitudes, $S_{\rm P_i}^{\rm lf}$ and $S_{\rm P_i}^{\rm hf}$, which serve as initial parameters for the numerical calculations. Here, the superscripts “lf” and “hf” denote the low- and high-frequency noise components, respectively. These estimates are obtained by taking into account the filter function of the Hahn-echo sequence. Additionally, the relative contributions from different gate electrodes are determined by their respective lever arms.

However, as shown in Fig.~\ref{fig:3}(c-d), the simulated $T_2^{\rm Echo}$ exhibits only a weak anisotropy as a function of the magnetic-field orientation $\theta_{\rm xy}$, in contrast to the pronounced anisotropic behavior observed experimentally. In addition, the simulated Hahn-echo coherence time $T_2^{\rm Echo}$ shows a significant deviation from the experimental results, whereas the simulated $T_2^{*}$ is in good agreement with the measurements. This difference is likely due to the fact that the Ramsey experiment probes free-induction decay and is therefore strongly influenced by low-frequency noise, which has a relatively large amplitude in our system. In contrast, the Hahn-echo sequence implements dynamical decoupling and partially suppresses the contribution from low-frequency noise. Previous experiment shows that the barrier gates also have a significant impact on the spin–electric susceptibility \cite{hendrickxSweetspotOperationGermanium2024}. However, noise contributions from these gates are not included in the present simulation. This omission arises because fast voltage pulses can only be applied to the plunger gates P1 and P2, preventing experimental access to the longitudinal spin–electric susceptibilities associated with the barrier gates B1–B3. Despite this limitation, the good agreement between the simulated and experimental values of $T_2^{*}$ demonstrates that the model captures the dominant low-frequency noise processes governing free-induction dephasing, while the remaining discrepancy in $T_2^{\rm Echo}$ highlights the importance of additional noise channels associated with the barrier gates that are not included in the present treatment.

\begin{figure}[htbp]
\includegraphics{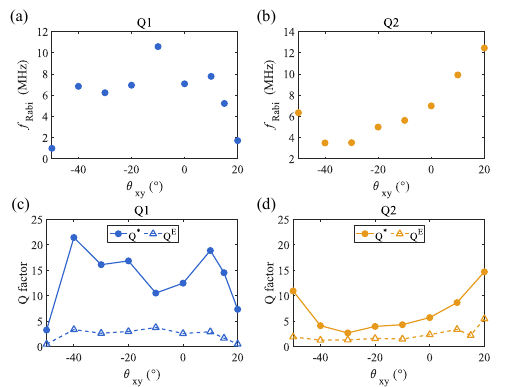}
\caption{\label{fig:6} (a–b) Rabi frequency $f_{\rm Rabi}$ of qubits Q1 and Q2 as functions of the in-plane magnetic-field angle $\theta_{\rm xy}$. For Q1, $f_{\rm Rabi}$ first increases and then decreases with $\theta_{\rm xy}$, opposite to the trends of LSES${_{\rm Q1}}$ and $T_2^{\rm Echo}$, indicating that strong spin–orbit coupling is a double-edged sword: it enhances qubit manipulation speed while simultaneously increasing spin–electric susceptibility. For Q2, $f_{\rm Rabi}$ increases monotonically with $\theta_{\rm xy}$, which is likely due to the microwave drive being applied only to gate P2, leading to a manipulation mechanism different from that of Q1. (c–d) Quality factors $Q^*$ and $Q^E$ of qubits Q1 and Q2 as functions of $\theta_{\rm xy}$. The quality factor $Q^*$ depends on both $T_2^*$ and $f_{\rm Rabi}$. Owing to the nearly constant $T_2^*$ over the measured range, $Q^*$ exhibits a trend similar to that of $f_{\rm Rabi}$. For qubit Q1, $Q^*$ remains approximately constant as $\theta_{\rm xy}$ varies, whereas for qubit Q2, $Q^*$ increases monotonically, following the enhancement of $f_{\rm Rabi}$. For Q1, the opposite angular dependences of $T_2^{\rm Echo}$ and $f_{\rm Rabi}$ result in no clear overall trend in $Q$. In contrast, for Q2, $Q$ is influenced by both $T_2^{\rm Echo}$ and $f_{\rm Rabi}$, showing a non-monotonic behavior that first decreases and then increases, following the trend of $T_2^{\rm Echo}$.}
\end{figure}

\section{\label{App-C}Magnetic field-dependent Rabi frequency and quality factor}

We characterize the Rabi frequency $f_{\rm Rabi}$ of Q1 and Q2. By combining the measured $f_{\rm Rabi}$ with the dephasing time $T_2^*$ and the Hahn-echo coherence time $T_2^{\rm Echo}$, the qubit quality factor can be defined as

\begin{equation}
Q^* = 2T_2^{*}f_{\rm Rabi} 
\end{equation}
\begin{equation}
Q^E = 2T_2^{\rm Echo}f_{\rm Rabi}
\end{equation}
The dependence of the Rabi frequency of the two qubits on the magnetic-field orientation is shown in Fig.\ref{fig:6}(a–b).

For qubit Q1, the Rabi frequency exhibits a non-monotonic dependence on the magnetic-field angle, first increasing and then decreasing. This behavior closely correlates with the variation of $T_2^{\rm Echo}$ shown in Fig.\ref{fig:3}(c), indicating that the decoherence of this qubit is predominantly limited by charge noise. In addition, the variation of $T_2^{\rm Echo}$ follows the same trend as the spin–electric susceptibility strength. Since the control of hole-spin qubits is fundamentally enabled by spin–orbit coupling, which simultaneously enhances the susceptibility to charge noise, regions with stronger spin–electric susceptibility generally correspond to shorter coherence times, while the control speed $f_{\rm Rabi}$ can be enhanced. This trade-off between control speed and coherence is clearly manifested in the correlated behavior of the spin–electric coupling strength, coherence time, and Rabi frequency of qubit Q1. The resulting quality factor $Q^*$ and $Q^E$ of qubit Q1, shown in Fig.\ref{fig:6}(c), does not display a pronounced monotonic dependence on the magnetic-field angle due to the opposite trends of $f_{\rm Rabi}$, $T_2^*$ and $T_2^{\rm Echo}$, with a maximum value of approximately 21.

In contrast, for qubit Q2, the Rabi frequency initially decreases with the magnetic-field angle and then increases rapidly, reaching a maximum value 12 MHz approximately four times larger than its minimum. Meanwhile, as shown in Fig.\ref{fig:3}(d), the Hahn-echo coherence time $T_2^{\rm Echo}$ of qubit Q2 remains nearly constant over a wide range of magnetic-field orientations, and even at angles where an enhancement is observed, the maximum value is only about twice the minimum. Consequently, the angular dependence of the quality factor $Q^*$ and $Q^E$ is primarily governed by $f_{\rm Rabi}$, exhibiting a trend that closely follows the variation of the Rabi frequency. The maximum quality factor of qubit Q2 is achieved at the magnetic-field orientation where $f_{\rm Rabi}$ is largest at $\theta_{\rm{xy}} \approx 20^{\circ}$, with a value of approximately 15, corresponding to a fivefold enhancement compared to its minimum value.

\nocite{*}

\bibliography{apssamp}

\end{document}